\documentclass[twocolumn,aps,showpacs,nofootinbib,prd]{revtex4-1}
\usepackage{amssymb}
\usepackage{amsmath}
\usepackage{graphics}
\usepackage{epsfig}
\usepackage[dvips]{color}

\newcommand{\be}{\begin{equation}}
\newcommand{\ee}{\end{equation}}
\newcommand{\bea}{\begin{eqnarray}}
\newcommand{\eea}{\end{eqnarray}}
\newcommand{\nn}{\nonumber\\}
\newcommand{\beas}{\begin{eqnarray*}}
\newcommand{\eeas}{\end{eqnarray*}}

\begin{document}

\title{Dimuon production from in-medium rho decays from QCD sum rules} 
\author{Alejandro Ayala$^1$, C. A. Dominguez$^2$, Luis Alberto Hern\'andez$^1$, M. Loewe$^{2,3}$, 
  Ana Julia Mizher$^1$} \address{$^1$Instituto de Ciencias
  Nucleares, Universidad Nacional Aut\'onoma de M\'exico, Apartado
  Postal 70-543, M\'exico Distrito Federal 04510,
  Mexico.\\ 
  $^2$Centre for Theoretical \& Mathematical Physics, University of Cape Town, Rondebosch 7700, South Africa.\\
  $^3$Facultad de F\1sica, Pontificia Universidad Cat\'olica de Chile, Casilla 306, Santiago 22, Chile}

\begin{abstract}

We compute the dimuon-excess invariant mass distribution at the $\rho$-meson peak in the context of relativistic heavy-ion collisions at SPS energies. The temperature, $T$, and chemical potential, $\mu$, dependent parameters describing the $\rho$-meson --its width, mass and leptonic decay constant-- are determined from finite energy QCD sum rules. Results show that the $\rho$-meson width increases whereas its mass and leptonic decay constant decrease near the (chemical potential-dependent) critical temperature $T_c(\mu)$ for chiral symmetry restoration/quark-gluon deconfinement. As a consequence, starting from $T_c(\mu)$, for a short lived cooling the main effect is a  broadening of the dimuon distribution. However, when the evolution brings the system to a lower freeze-out temperature, with the $\rho$-meson parameters approaching their vacuum values, the dimuon distribution shows a broad peak centered at the vacuum $\rho$-meson mass. For even lower freeze-out temperatures the peak becomes less prominent since the thermal phase space factor suppresses the distribution for larger values of the invariant dimuon mass, given that the average temperature is smaller. The dimuon distribution exhibits a non-trivial behavior with $\mu$. For small $\mu$ values the distribution broadens with increasing $\mu$ becoming a bit steeper at low invariant masses for larger values of $\mu$. Our results are in very good agreement with data from the NA60 Collaboration. 
\end{abstract}

\maketitle

Electromagnetic probes reveal the entire thermal evolution of a heavy-ion collision, as they are emitted continuously from the early quark-gluon plasma phase, through to the late hadronic phase up to freeze out. After being produced, these probes escape from the medium without further interaction, as their mean free path is larger than the size of the fireball. Low mass dileptons are one of these probes and their invariant-mass spectrum is a direct measurement of the in-medium hadronic spectral function in the vector channel. For invariant masses below 1 GeV, the spectrum is dominated by the $\rho$-meson whose short lifetime and large coupling to pions and muons make it an ideal test particle to sample in-medium changes of the parameters describing it, such as its mass, its width and its leptonic decay constant. The temperature and density dependence of these parameters provides information on the phase transitions to quark-gluon deconfinement and chiral symmetry restoration.

In the past few years, high quality data provided by the CERN NA60 Collaboration~\cite{NA60-1,NA60-2} has allowed to settle a long standing controversy regarding the origin of the dilepton excess~\cite{CERES} below the $\rho$-meson peak at SPS energies. This excess was originally attributed either to a mass decrease or to an increase in the width of the $\rho$-meson with increasing temperature. The NA60 data shows that when the known sources are subtracted from the spectral function, a clear peak at the vacuum $\rho$ mass is observed at all centralities. The peak broadens for the most central collisions but remains centered at its vacuum mass value. The total dimuon yield also increases with centrality. This result is in line with current ideas about the nature of the deconfinement/chiral symmetry restoration transition. Since the mass is only related to the position of the real part of the hadron propagator in the complex squared energy plane, its temperature behavior provides no relevant information on deconfinement. The relevant quantity is, instead, the hadronic width related to the imaginary part of the propagator. A width which increases with increasing $T$ does indicate deconfinement, as the spectral function eventually becomes smooth and is accounted for by perturbative QCD.

The leading theoretical descriptions of the $\rho$-meson broadening at SPS energies are based on hadronic many-body or transport calculations, invoking the idea that the $\rho$-meson scatters and thus melts not only at finite temperature, but also in a baryon-rich environment~\cite{Rapp,Hees-Rapp,Cassing,Weil,Rapp2}. These rather successful descriptions have allowed the interpretation of a wealth of fixed target data. In principle, this approach could also be expected to describe data at RHIC and LHC. In fact the net density remains comparable to SPS energies, in spite of the baryon density becoming smaller. However, results from these approaches applied to collider energies are not as successful. In fact, the PHENIX Collaboration has found a large enhancement in central Au + Au at $\sqrt{s_{NN}}= 200$ GeV~\cite{PHENIX} which cannot be described by the in-medium hadronic effects invoked at SPS energies, albeit the description works well for lower centralities and lower energies. It then becomes imperative to explore alternative descriptions which could be valid in both energy regions, and shed some light on the origin of this enhancement.

In this paper we make use of recent results from QCD sum rules (QCDSR) at finite $T$ in the vector channel~\cite{nos}, and extend them to finite $\mu$. This method provides information on the thermal and density behavior of the $\rho$-meson parameters, i.e. its width, coupling, and mass, which are needed to compute the dimuon production rate in heavy-ion collisions around the $\rho$-peak.

The starting point is the light-quark vector current correlator, that at $T = 0$ can be written as
\bea
   \Pi_{\mu\nu}&=&i\int d^4xe^{iqx}\langle 0 |{\mathcal{T}}[V_\mu (x) V^\dagger_\nu(0)]|0\rangle\nn
   &=&(-g_{\mu\nu}q^2 + q_\mu q_\nu)\Pi (q^2),
\label{correlatorT=0}
\eea
where $V_\mu=(1/2)[:\bar{u}(x)\gamma_\mu u(x) - \bar{d}(x)\gamma_\mu d(x):]$ is the (electric-charge neutral) conserved vector current in the chiral limit, and $q_\mu =(\omega, {\bf q})$ is the four-momentum carried by the current. The function $\Pi (q^2)$ in pQCD is normalized as ${\mbox{Im}} (q^2)=1/(8\pi )[1+{\mathcal{O}}(\alpha_s(q^2))]$. Finite energy QCDSR (FESR) rest on two pillars~\cite{Colangelo}, (i) the
operator product expansion (OPE) of current correlators at short distances beyond perturbation theory, and (ii)
Cauchy's theorem in the complex squared energy $s$-plane, which relates the (hadronic) discontinuity across the cut on the real positive semi-axis with the integral around a contour of radius $|s_0|$ where the OPE is expected to be valid. The latter is usually referred to as quark-hadron duality. This leads to the FESR
\bea
   (-1)^{N-1}C_{2N}\langle O_{2N}\rangle &=& 8\pi^2\left[
   \int_0^{s_0}dss^{N-1}\frac{1}{\pi}{\mbox{Im}}\Pi^{\mbox{\small{HAD}}}(s)\right.\nn
   &-& \left.
   \int_0^{s_0}dss^{N-1}\frac{1}{\pi}{\mbox{Im}}\Pi^{\mbox{\small{pQCD}}}(s)\right],
\label{FESR}
\eea
where $N =1,\ 2,\ldots$, and the leading order vacuum condensates in the chiral limit are the dimension $d = 4$ gluon condensate $C_4\langle O_4\rangle = (\pi/3)\langle\alpha_sG^2\rangle$, and the dimension $d=6$ four-quark condensate, $C_6\langle O_6\rangle$. The extension of the QCDSR to finite temperature was proposed in Ref.~\cite{Bochkarev} and justified in quantum field theory models in Ref.~\cite{Dominguez-Loewe}. The hadronic spectral function is well approximated by the Breit-Wigner form
\bea
   \frac{1}{\pi}{\mbox{Im}}\Pi_0^{\mbox{\small{HAD}}}(s)= \frac{1}{\pi}\frac{1}{f_\rho^2}\frac{M_\rho^3\Gamma_\rho}{(s-M_\rho^2)^2+M_\rho^2\Gamma_\rho^2},
\label{BW}
\eea
where $f_\rho$, $M_\rho$ and $\Gamma_\rho$ are the $\rho$-meson leptonic decay constant, mass and width, respectively. 
At finite $T$ all hadronic parameters become $T$-dependent, and so do the condensates. In the thermal perturbative QCD sector, only the leading one-loop contributions can be taken into account, since the problem of the appearance of two scales, i.e. the short distance QCD scale and the critical temperature $T_c$, remains unsolved (QCDSR evolve from $T = 0$ up to $T_c$, i.e. a region where ordinary thermal perturbation theory is not valid). At finite $T$ and $\mu = 0$ the results of Ref.~\cite{nos} for the $\rho$-meson width and mass are
\bea
   \Gamma_\rho (T)&=&\Gamma_\rho (0)\left[1 - (T/T_c)^3\right]^{-1},\nn
   M_\rho (T)&=&M_\rho (0)\left[1 - (T/T_M^*)^{10}\right],
\label{inputs}
\eea
with $\Gamma_\rho (0)=0.145$ GeV and $M_\rho (0)=0.776$ GeV. The solutions can be written as functions of the scaled temperature $T/T_c$. Here we chose $T_c=0.197$ GeV. Also, $T_M^*=0.222$ GeV. The other quantities are:
$
C_6\langle O_6\rangle= C_6\langle O_6\rangle (0)[1-(T/T_q^*)^8]
$,
$   
   s_0(T)=s_0(0)[1-0.5667(T/T_c)^{11.38} -4.347(T/T_c)^{68.41}]
$,
$
   C_4\langle O_4\rangle (T)=C_4\langle O_4\rangle (0)[1-1.65(T/T_c)^{8.73}+0.04967(T/T_c)^{0.72}]
$,
$
   f_\rho(T)=f_\rho(0)[1-0.3901(T/T_c)^{10.75}+0.04155(T/T_c)^{1.27}]
$,
with $s_0(0)=1.7298$ GeV$^2$, $C_4\langle O_4\rangle (0)=0.412561$ GeV$^4$, $C_6\langle O_6\rangle (0)=-0.951667$ GeV$^6$, $T_q^*=0.187$ GeV and $f_\rho(0)=5$. 

\begin{widetext}
\begin{figure*}[t!]
\begin{center}
\begin{tabular}{ccc}
\includegraphics[scale=0.57]{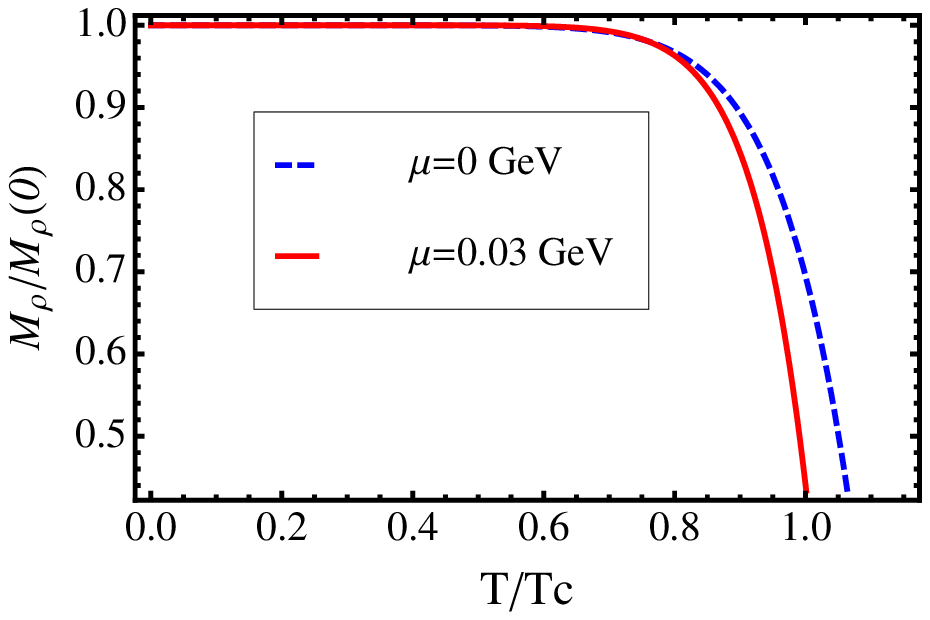} &
\includegraphics[scale=0.57]{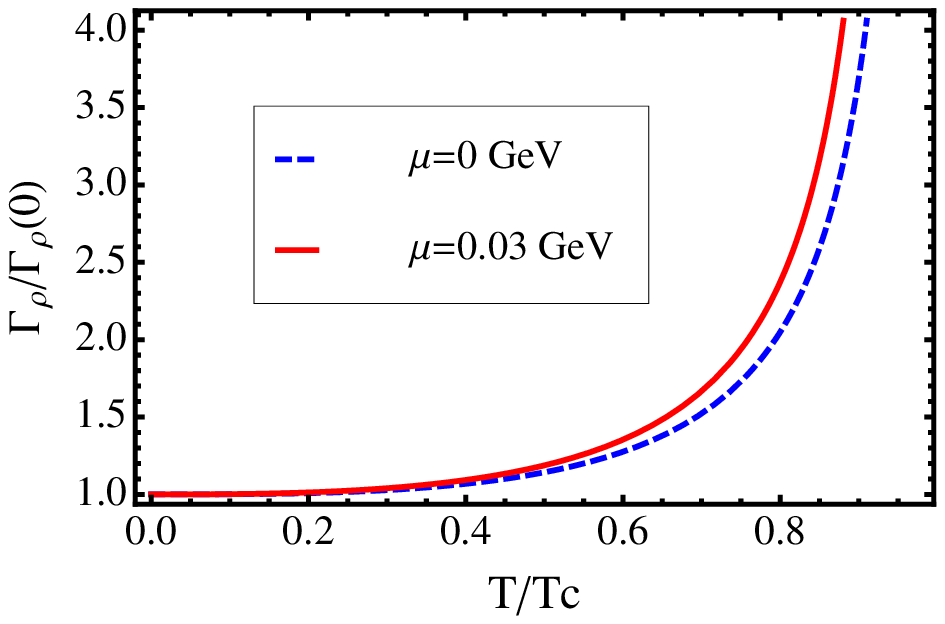} &
\includegraphics[scale=0.57]{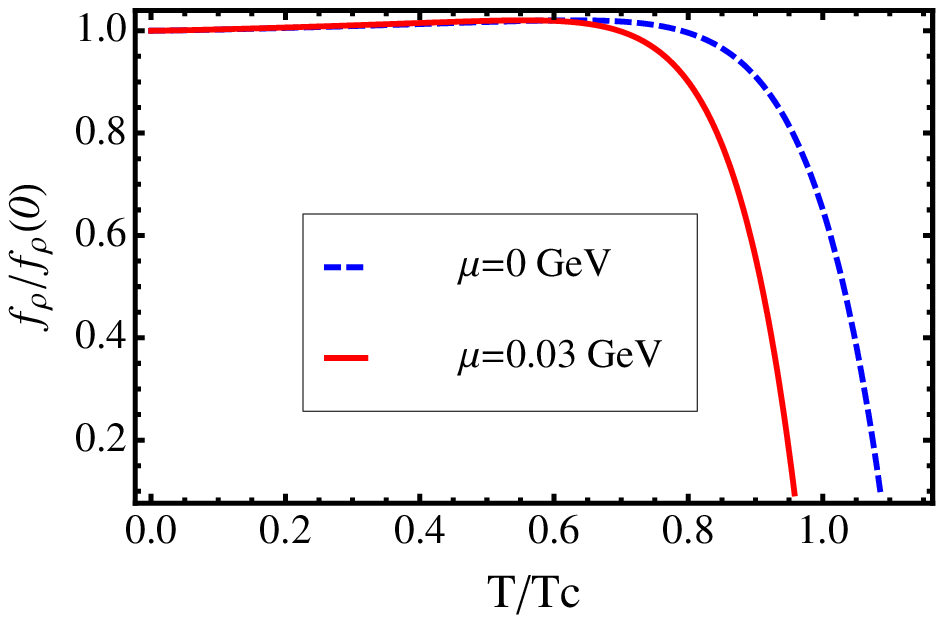}
\\ \ \ \ (a) &\ \ \ (b) &\ \ \ (c)
\end{tabular}
\end{center}
\caption{Temperature behavior of the $\rho$ parameters for $\mu=0,\ 0.03$ GeV. (a) $M_\rho$, (b) $\Gamma_\rho$ and (c) $f_\rho$.}
\label{fig1}
\end{figure*}
\end{widetext}

In order to extend this analysis to finite chemical potential we first incorporate the $\mu$-dependence into the second term on the right-hand side of Eq.~(\ref{FESR}), which involves a quark loop. This modifies the corresponding Fermi-Dirac distribution, splitting it into particle--antiparticle contributions. Next, we incorporate the $\mu$-dependence of the critical temperature $T_c$. For this, we resort to the findings in Ref.~\cite{morelia} that provide, using a Schwinger-Dyson approach, a parametrization for the crossover transition line between chiral symmetry restored and broken phases, valid for small values of $\mu$,
\bea
   T_c(\mu)=T_c(\mu=0) - 0.218\mu - 0.139\mu^2.
\label{tcvsmu}
\eea
In order to solve the FESR we follow the procedure of Ref.~\cite{nos} and use as inputs the parameters $s_0(T,\mu)$, $f_\rho(T,\mu)$ and $C_4\langle O_4\rangle(T,\mu)$, together with Eq.~(\ref{tcvsmu}), and obtain $M_\rho(T,\mu)$, $\Gamma_\rho(T,\mu)$ and $C_6\langle O_6\rangle(T,\mu)$. For $\mu=0.03$ GeV, we find: 
$   
   M_\rho(T,\mu)=M_\rho(0)[1-0.5597(T/T_c)^{12.18}]
$,
$
   \Gamma_\rho(T,\mu)=\Gamma_\rho(0)[1-1.0717(T/T_c)^{2.763}]^{-1}
$,
$
   f_\rho(T,\mu)=f_\rho(0)[1-0.3901(T/T_c(\mu))^{10.75}+0.04155(T/T_c(\mu))^{1.27}]
$.

Figure~\ref{fig1} shows $M_\rho(T,\mu)$, $\Gamma_\rho(T,\mu)$ and $f_\rho(T,\mu)$ as functions of $T/T_c$ normalized to their values at $T=\mu=0$ for $\mu=0$ and $\mu=0.03$ GeV (corresponding to a baryon chemical potential $\mu_B=3\mu=0.09$ GeV). 

With these solutions we proceed to compute the dimuon thermal rate in the hadronic phase originating from $\rho$--decays. We consider processes where pions annihilate into $\rho$'s which in turn decay into dimuons by means of vector dominance. The rate is given by
\bea
   \frac{dN}{d^4xd^4K}&=&\frac{\alpha^2}{48\pi^4}
   \left(1+\frac{2m^2}{M^2}\right)\left(1-\frac{4m_\pi^2}{M^2}\right)
   \sqrt{1-\frac{4m^2}{M^2}}\nn
   &\times&
   e^{-K_0/T}{\mathcal{R}}(K,T){\mbox{Im}}\Pi_0^{\mbox{\tiny{res}}}(M^2),
\label{rate}
\eea
where $N$ is the number of muon pairs per unit of infinitesimal space-time and energy-momentum volume, with $x^\mu =(t,{\mathbf{x}})$ the space-time coordinate and $K^\mu=(K_0,{\mathbf{K}})$ ($K=|{\mathbf{K}}|$) the four-momentum of the muon pairs, $\alpha$ is the electromagnetic coupling, $m$ is the muon mass, $m_\pi$ is the pion mass, $M$ is the dimuon invariant mass, ${\mbox{Im}}\Pi_0^{\mbox{\tiny{res}}}(M^2)$ is given by Eq.~(\ref{BW}) and
\bea
   {\mathcal{R}}&=&\frac{T/K}{1-e^{-K_0/T}}\nn
   &\times&\ln\left[
   \left(\frac{e^{-E_{\mbox{\tiny{max}}}/T}-1}{e^{-E_{\mbox{\tiny{min}}}/T}-1}\right)
   \left(\frac{e^{E_{\mbox{\tiny{min}}}/T}-e^{-K_0/T}}{e^{E_{\mbox{\tiny{max}}}/T}-e^{-K_0/T}}\right)
   \right],
\label{funcR}
\eea
with
\bea
   E_{\mbox{\tiny{max}}}&=&\frac{1}{2}\left[K_0 + K\sqrt{1-4m_\pi^2/M^2}\right]\nn
   E_{\mbox{\tiny{min}}}&=&\frac{1}{2}\left[K_0 - K\sqrt{1-4m_\pi^2/M^2}\right].
\label{eminmax}
\eea
\begin{figure}[t]
\centering\includegraphics[width=\columnwidth]{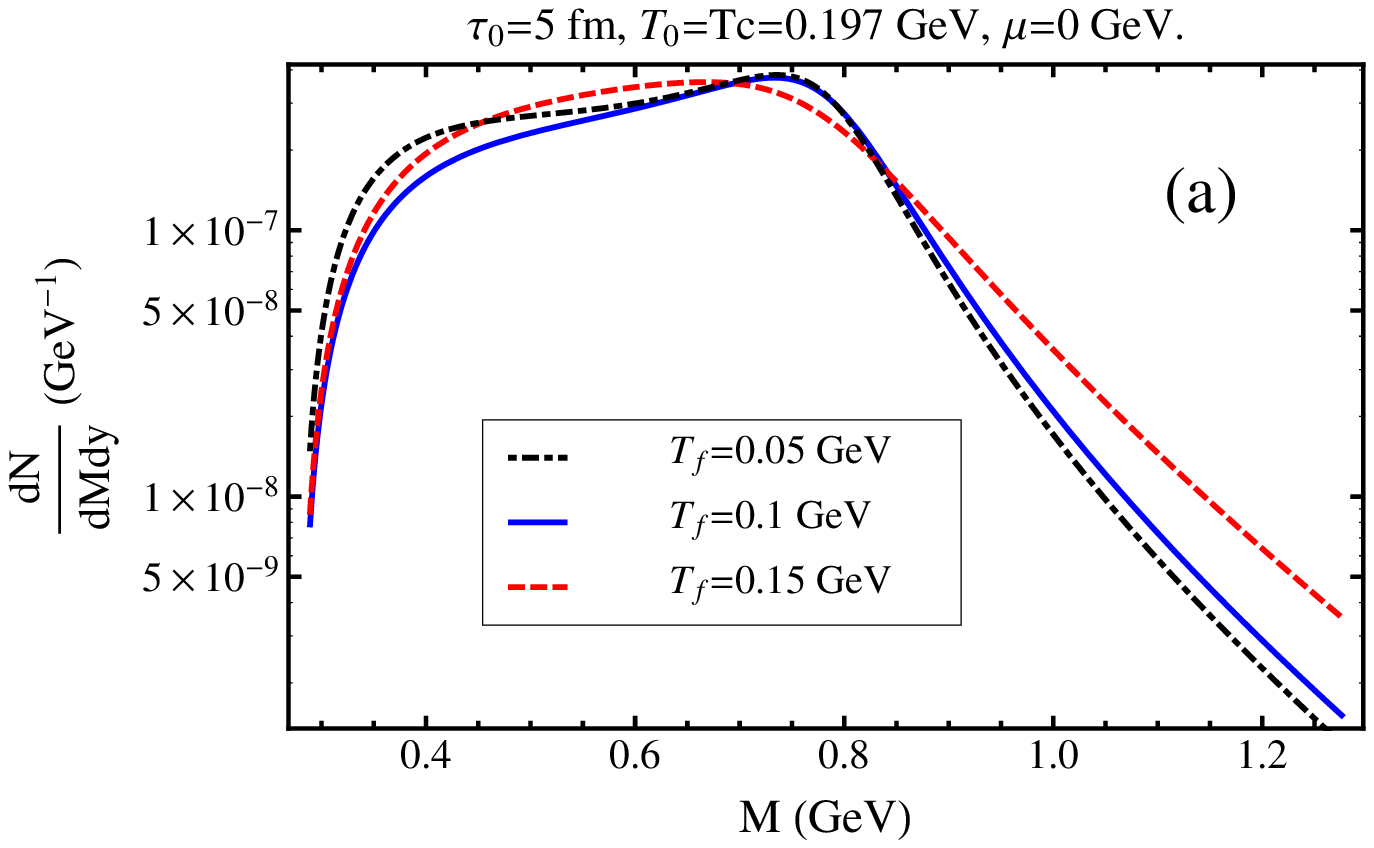}\\
\centering\includegraphics[width=\columnwidth]{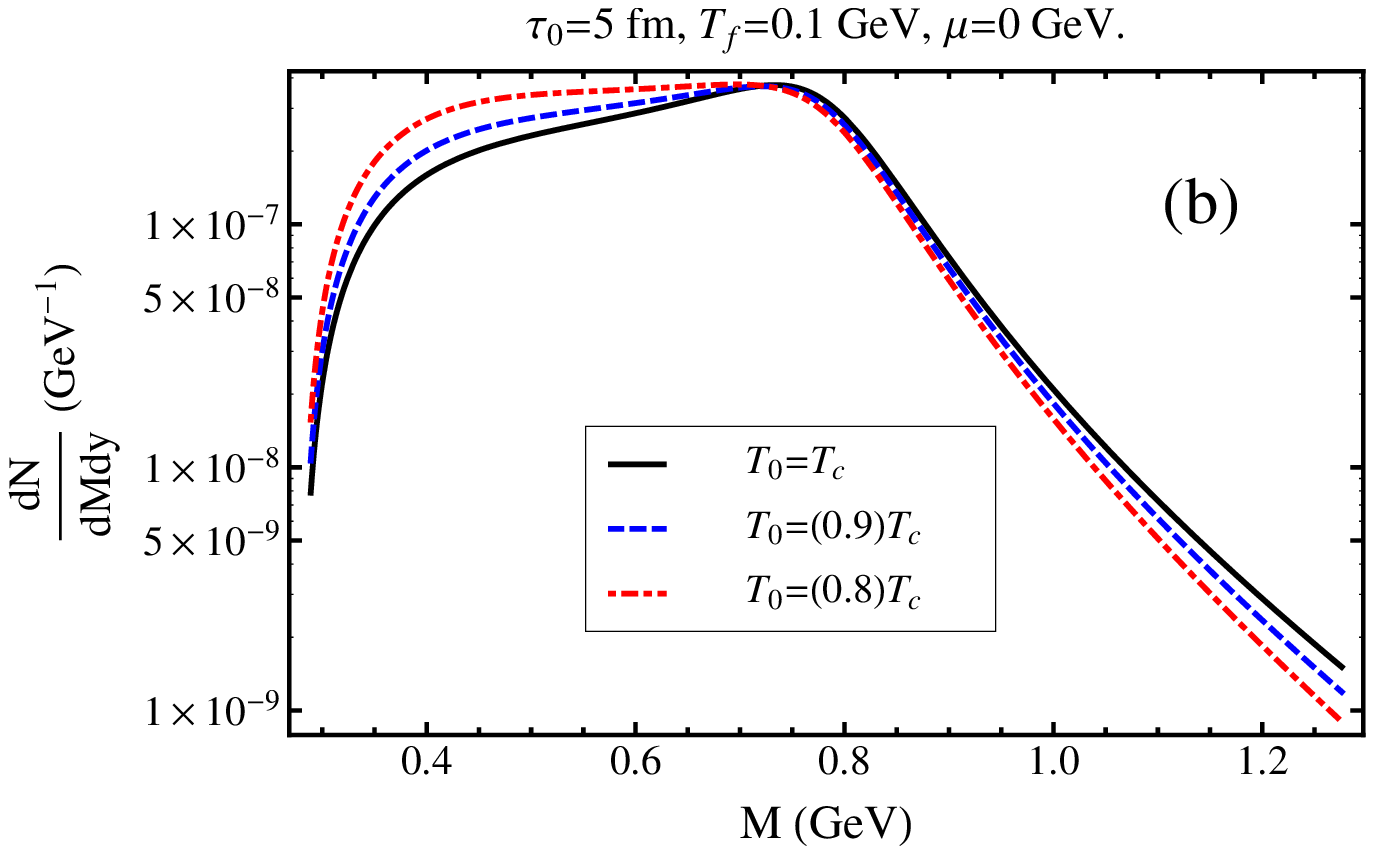}\\
\centering\includegraphics[width=\columnwidth]{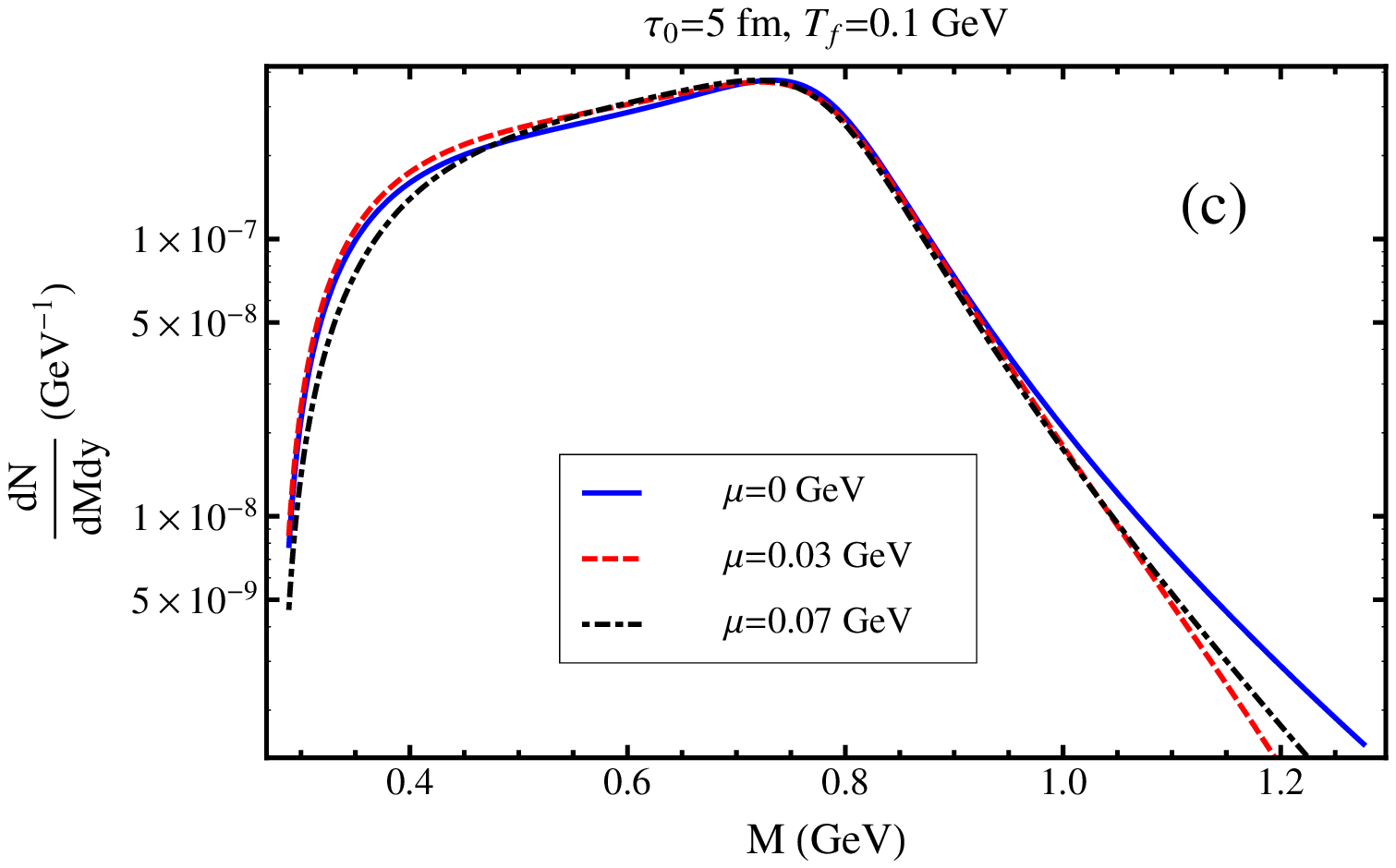} 
\caption{Invariant dimuon mass distribution for different values of (a) $T_f$, (b) $T_0$ and (c) $\mu$.}
\label{fig2}
\end{figure}
In order to integrate Eq.~(\ref{rate}) we use
\bea
   d^4K&=&\frac{1}{2}dM^2d^2K_\perp dy\nn
   d^4x&=&\tau d\tau d\eta d^2x_\perp,
\label{phasespace}
\eea
where $y$ and $\eta$ are the momentum-space and coordinate-space rapidities, respectively and $\tau=\sqrt{t^2-z^2}$. To relate the temperature change to the time evolution of the system, we neglect a possible small transverse expansion, assume it entirely longitudinal, and use the cooling law
\bea
   T=T_0\left(\frac{\tau_0}{\tau}\right)^{v_s^2},
\label{cooling}
\eea
where $v_s^2=1/3$ is the square of the sound velocity for an ideal hadron gas. The starting initial proper-time is taken as $\tau_0=5$ fm with changes in this value amounting to a rescaling of the distribution but not affecting its shape. The evolution is taken down to a freeze-out temperature $T_f$. From Eq.~(\ref{cooling}), a change in $\tau_0$ allows to change the chosen value of $T_0$, namely of $T_c$ and the shape of the distribution is maintained provided the ratio $T_f/T_0$ is preserved. Also, we consider perfect correlation between $\eta$ and $y$ ($\eta=y$). The invariant mass distribution becomes
\bea
   \!\!\!\frac{dN}{dMdy}=\Delta y M\int_{\tau_0}^{\tau_f}\tau d\tau\int d^2K_\perp \int d^2x_\perp
   \frac{dN}{d^4xd^4K}.
   \label{invmass}
\eea
\begin{figure}[t!]
\centering\includegraphics[width=\columnwidth]{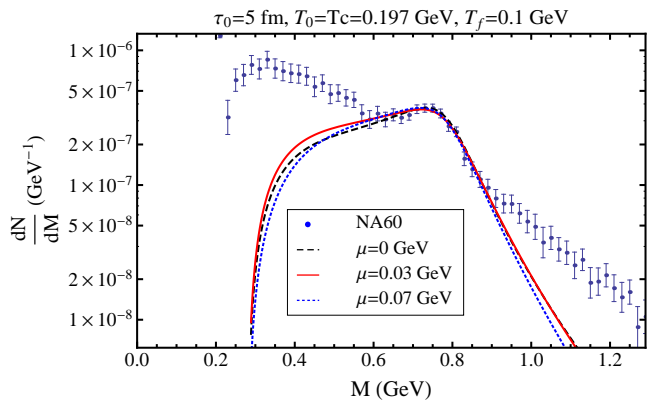}
\caption{Invariant dimuon mass distribution compared to NA60 data. The calculation is performed for three different values of $\mu=0, \  0.03, \ 0.07$ GeV.}
\label{fig3}
\end{figure}

Figure~\ref{fig2} shows examples of the invariant mass dimuon distribution $dN/dMdy$, normalized to the NA60 data around the $\rho$ peak. Figure~\ref{fig2}(a) shows the invariant mass distribution for $\mu=0$, an initial temperature $T_0=T_c=0.197$ GeV and three freeze-out temperatures, $T_f=0.05,\ 0.1,\ 0.15$ GeV. Notice that for a larger $T_f$ the dominant effect is the broadening of the $\rho$ peak which can be traced back to the broadening of $\Gamma$ for temperatures closer to $T_c$.  For a smaller freeze-out temperature, and therefore a larger evolution time, the $\rho$ peak is more clearly visible since the cooler the system, the more vacuum-like the $\rho$ parameters become. When the freeze-out temperature becomes even smaller, the $\rho$ peak becomes less prominent since the dominant effect is the thermal phase space factor which suppresses the distribution for larger values of $M$, given that the average temperature is smaller. In this sense, the distribution is also sensitive to the value of $v_s$ and therefore, more generally, to the equation of state. Figure~\ref{fig2}(b) shows the invariant mass distribution for $T_f=0.1$ GeV, $\mu=0$ and three initial temperatures $T_0=T_c,\ 0.9T_c,\ 0.8T_c$. Notice that the smaller the initial temperature, the flatter the distribution. This can once again be understood as an effect due to the smaller average temperature that suppresses the distribution for larger values of $M$. Figure~\ref{fig2}(c) shows the invariant mass distribution for $T_f=0.1$ GeV, $T_0=T_c$ and three values of $\mu=0,\ 0.03,\ 0.07$ GeV. Notice that the distribution shows a non-trivial behavior with $\mu$. For small $\mu$ the distribution is enhanced at small values of $M$ at the expense of being suppressed for large values of $M$. However with increasing $\mu$ the distribution becomes a bit steeper at small $M$. This behavior could be an interesting clue when applying these ideas to describing the dimuon excess radiation at collider energies where the baryon density is smaller.

Figure~\ref{fig3} shows the results for $dN/dM$ compared to the NA60 data around the $\rho$ peak for $T_0=T_c=0.197$ GeV, $T_f=0.1$ GeV and for three values $\mu=0,\ 0.03,\ 0.07$ GeV. The given values of $\mu$ are here to be regarded as average values during the evolution of the collision. The theoretical results provide a very a good description of the data around the $\rho$ peak. 

The parameters used in the calculation are not unique, however we emphasize that an important quantity for the shape of the distribution is the ratio $T_0/T_f$, which in our case is about a factor 2. The shape of the distribution can be preserved even if the above ratio changes by changing $v_s$ which in turn implies a sensitivity to the equation of state. Notice the threshold from twice the pion mass which is due to the fact that the only process we are considering is the pion annihilation channel into $\rho$'s that later decay into dimuons. When the dimuon pair has an invariant mass lower than twice the pion mass, this process cannot happen. This indicates that for the description of the data at lower and higher invariant masses, additional processes need to be accounted for~\cite{Hees-Rapp}. For instance, for lower invariant masses, one needs to consider the scattering of pions and $\rho$ mesons. This process does not have a threshold. Moreover if the scattering happens with off-shell $\rho$'s in the initial state, then the temperature and density evolution of the $\rho$ parameters becomes also relevant.
For higher invariant masses other vector resonances as well as the quark-gluon plasma contributions to the excess radiation need to be considered. In addition, and at the $\rho$ peak itself, freeze-out and primordial $\rho$ mesons, and a transverse expansion could contribute to further shape the peak. 

In conclusion, we have shown that the QCD FESR are a powerful tool to compute the $\rho$-meson parameters at finite temperature and chemical potential, thus providing a complementary approach to many-body descriptions of in-medium hadron properties. We have used these parameters to compute the dimuon-excess invariant mass distribution for temperatures and densities relevant to SPS energies. The description is in very good agreement with NA60 data in the region of the $\rho$-peak. Further studies including the effects of transverse expansion as well as other contributions to the dimuon radiation near the $\rho$ peak are being prepared and will be reported elsewhere.

\section*{Acknowledgments}

A.A. is in debt to Y. Zhang for helpful discussions, to R. Rapp for useful comments and to H. Specht, S. Damjanovic and E. Scomparin for discussing and making available the NA60 data. This work has been supported in part by DGAPA-UNAM under grant PAPIIT-IN103811, CONACyT-M\'exico under grant 128534, FONDECyT (Chile) under grants 1130056 and 1120770, NRF (South Africa), and the University of Cape Town URC.

\end{document}